\documentclass{PoS}

\usepackage{amsmath}
\usepackage{url}
\usepackage{subfig}

\title{Probing the high-$x$ content of the nuclei in the fixed-target mode at the LHC}

\ShortTitle{Probing the high-$x$ content of the nuclei in the fixed-target mode at the LHC}

\author{\speaker{A.~Kusina}$^a$, C.~Hadjidakis$^b$, D.~Kiko{\l}a$^c$, J.P.~Lansberg$^b$,
        L.~Massacrier$^b$, M.G.~Echevarria$^d$, I.~Schienbein$^e$, J.~Seixas$^{f,g}$, 
        H.S.~Shao$^h$, A.~Signori$^i$, B.~Trzeciak$^j$, S.J.~Brodsky$^k$, G.~Cavoto$^l$,
        C.~Da~Silva$^m$, F.~Donato$^n$, E.G.~Ferreiro$^{o,p}$, I.~H\v{r}ivn\'{a}\v{c}ov\'{a}$^b$,
        A.~Klein$^m$, A.~Kurepin$^r$, C.~Lorc\'e$^s$, F.~Lyonnet$^t$, Y.~Makdisi$^u$,
        S.~Porteboeuf$^w$, C.~Quintans$^g$, A.~Rakotozafindrabe$^x$, P.~Robbe$^y$,
        W.~Scandale$^z$, N.~Topilskaya$^r$, A.~Uras$^{aa}$, J.~Wagner$^{ab}$, N.~Yamanaka$^b$,
        Z.~Yang$^{ac}$, A.~Zelenski$^u$\\
        {\scriptsize
        \llap{$^a$}Institute of Nuclear Physics Polish Academy of Sciences, PL-31342 Krakow, Poland\\
        \llap{$^b$}IPNO, CNRS-IN2P3, Univ. Paris-Sud, Universit\'e Paris-Saclay, 91406 Orsay Cedex, France\\
        \llap{$^c$}Faculty of Physics, Warsaw University of Technology, ul. Koszykowa 75, 00-662 Warsaw, Poland\\
        \llap{$^d$}Istituto Nazionale di Fisica Nucleare, Sezione di Pavia, via Bassi 6, 27100 Pavia, Italy\\
        \llap{$^e$}Laboratoire de Physique Subatomique et de Cosmologie, Universit\'e Grenoble Alpes,
                   CNRS/IN2P3, 53 Avenue des Martyrs, F-38026 Grenoble, France\\
        \llap{$^f$}Dep. Fisica, Instituto Superior Tecnico, Av. Rovisco Pais 1, 1049-001 Lisboa, Portugal\\
        \llap{$^g$}LIP, Av. Prof. Gama Pinto, 2, 1649-003 Lisboa,Portugal\\
        \llap{$^h$}LPTHE, UMR 7589, Sorbonne University\'e et CNRS, 4 place Jussieu, 75252 Paris, France\\
        \llap{$^i$}Physics Division, Argonne National Laboratory, Lemont, IL 60439, USA\\
        \llap{$^j$}Institute for Subatomic Physics, Utrecht University, Utrecht, The Netherlands\\
        \llap{$^k$}SLAC National Accelerator Laboratory, Stanford University, Menlo Park, CA 94025, USA\\
        \llap{$^l$}``Sapienza" Universit\`a di Roma, Dipartimento di Fisica \&
                   INFN, Sez. di Roma, P.le A. Moro 2, 00185 Roma, Italy\\
        \llap{$^m$}P-25, Los Alamos National Laboratory, Los Alamos, NM 87545, USA\\
        \llap{$^n$}Turin University, Department of Physics, and INFN, Sezione of Turin, Turin, Italy\\
        \llap{$^o$}Dept. de F{\'\i}sica de Part{\'\i}culas \& IGFAE, Universidade de Santiago de Compostela,
                   15782 Santiago de Compostela, Spain\\
        \llap{$^p$}Laboratoire Leprince-Ringuet, Ecole polytechnique, CNRS/IN2P3, Universit\'e Paris-Saclay, Palaiseau, France\\
        \llap{$^r$}Institute for Nuclear Research, Moscow, Russia\\
        \llap{$^s$}CPHT, \'Ecole Polytechnique, CNRS,  91128 Palaiseau, France\\
        \llap{$^t$}Southern Methodist University, Dallas, TX 75275, USA\\
        \llap{$^u$}Brookhaven National Laboratory, Collider Accelerator Department\\
        \llap{$^w$}Universit\'e Clermont Auvergne, CNRS/IN2P3, LPC, F-63000 Clermont-Ferrand, France\\
        \llap{$^x$}IRFU/DPhN, CEA Saclay, 91191 Gif-sur-Yvette Cedex, France\\
        \llap{$^y$}LAL, Universit\'e Paris-Sud, CNRS/IN2P3, Orsay, France\\
        \llap{$^z$}CERN, European Organization for Nuclear Research, 1211 Geneva 23, Switzerland\\
        \llap{$^{aa}$}IPNL, Universit\'e Claude Bernard Lyon-I and CNRS-IN2P3, Villeurbanne, France\\
        \llap{$^{ab}$}National Centre for Nuclear Research (NCBJ), Ho\.{z}a 69, 00-681, Warsaw, Poland\\
        \llap{$^{ac}$}Center for High Energy Physics, Department of Engineering Physics, Tsinghua University, Beijing, China\\}
        }

\vspace*{-1.2cm}        
        
\abstract{Using the LHCb and ALICE detectors in the fixed-target mode at the LHC offers
unprecedented possibilities to study the quark, gluon and heavy-quark content of the proton and nuclei
in the poorly known region of the high-momentum fractions. We review our projections for
studies of Drell-Yan, charm, beauty and quarkonium production with both detector
set-ups used with various nuclear targets and the LHC proton beams. Based on this, we
show the expected improvement in the determination of the quark, charm and gluon proton and nuclear
PDFs as well as discuss the implication for a better understanding of the cold-nuclear-matter
effects in hard-probe production in proton-nucleus collisions.}

\FullConference{International Conference on Hard and Electromagnetic Probes of High-Energy Nuclear Collisions\\
		30 September - 5 October 2018\\
		Aix-Les-Bains, Savoie, France}

\begin{document}

\section{Introduction}
\label{sec:intro}
In order to complement the current program of the LHC experiments, the LHC beam can be used to 
run in a fixed target mode. A fixed target experiment has a number of advantages that were
discussed in detail within the scope of the AFTER@LHC proposal~\cite{Brodsky:2012vg,Hadjidakis:2018ifr},
here we just mention the main ones:
(i) access to the far backward c.m.s. region,
(ii) possibility of using various targets including $^3$He, 
(iii) target polarisation giving access to measurements of single spin asymmetry at large momentum fractions.
For the details on the design and technical implementation of the experiment we refer the reader to the full
report~\cite{Hadjidakis:2018ifr} we just want to highlight here that scenarios using either a new detector
or one of the existing detectors (LHCb or ALICE) were consider; in both cases allowing for running
alongside the current collider program of the LHC. 
A three main physics subjects have been studied 
(i) the high fractional momentum ($x$) frontier in nucleons and nuclei,
(ii) spin content of the nucleons,
and (iii) heavy-ion physics.
In this contribution we concentrate on the first one -- the high-$x$ frontier -- for the results
on the remaining topics we refer the reader to the report~\cite{Hadjidakis:2018ifr}
and to other contributions from the Hard Probes 2018
conference~\cite{Yamanaka:HardProbes2018,Hadjidakis:HardProbes2018,Uras:HardProbes2018,Lansberg:HardProbes2018}.

\begin{figure}[hbt!]
\centering
\includegraphics[width=0.85\textwidth]{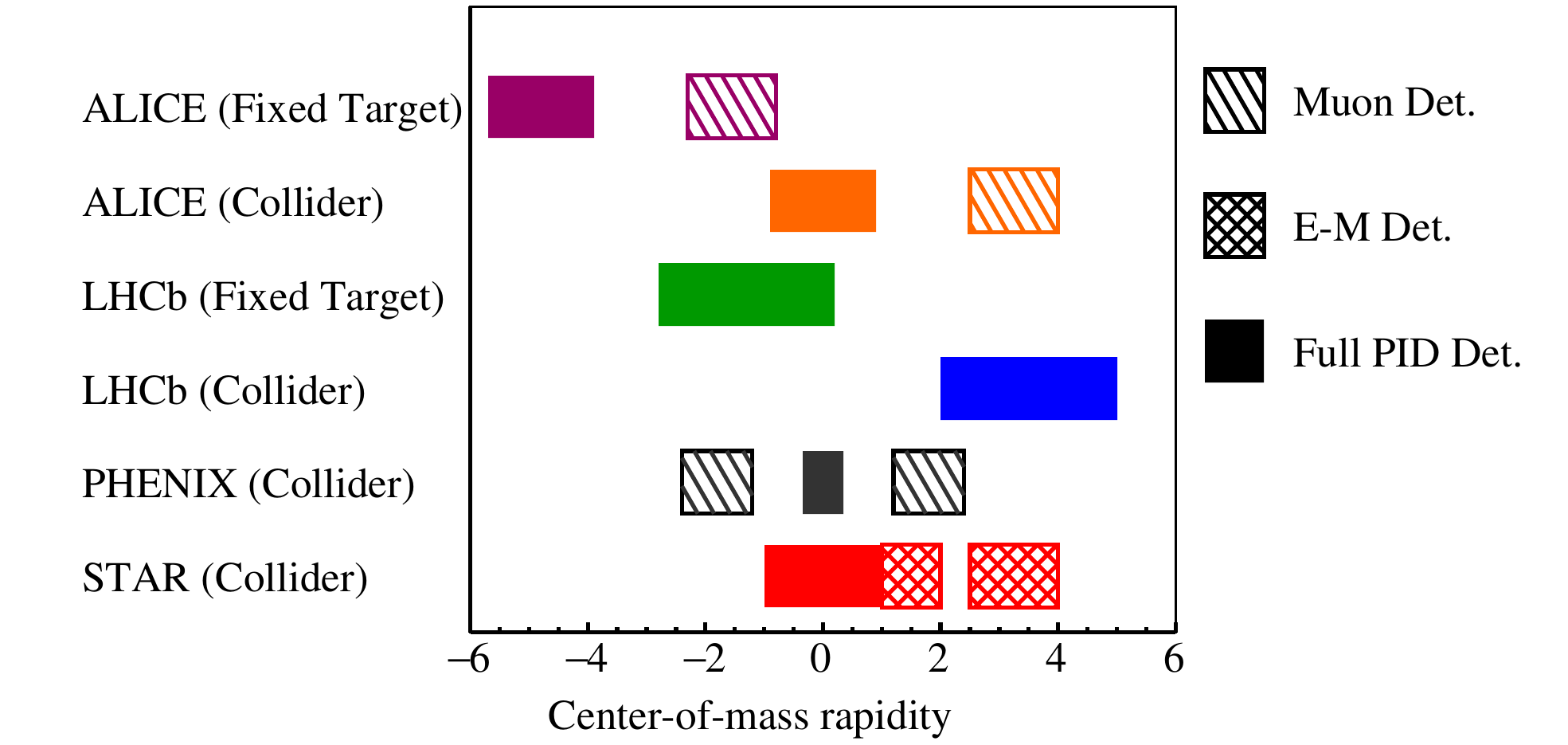} 
\caption{\small Comparison of the kinematic coverages of the ALICE and LHCb detectors at the LHC and
the STAR and PHENIX detectors at RHIC. For ALICE and LHCb, the acceptance is shown in the collider
and the fixed-target modes with a target position at the nominal Interaction Point (IP) for a
7 TeV proton beam. The "Full PID Det." label indicates detector with particle identification
capabilities, "E-M Det." - an electromagnetic calorimeter, "Muon Det." - a muon detector.
Figure from~\cite{Hadjidakis:2018ifr}.} 
\label{fig:acceptanceComparison-LHC-RHIC}
\end{figure}

Scenarios using a 7 TeV proton beam and a 2.76 TeV lead beam were considered. In the fixed target mode
these correspond to the nucleon-nucleon center of mass energy (c.m.s.) of $\sqrt{s_{NN}}=115$ GeV in the
proton case and $\sqrt{s_{NN}}=72$ GeV in the lead case.
The high-energy of the beams translates into a big pseudorapidity shift between the laboratory and c.m.s. frames
which is 4.8 for the proton beam and 4.3 for the lead beam. This means that we obtain a backward detector
covering $y_{\mathrm{c.m.s.}}<0$ and negative Feynman $x$ ($x_B$). The acceptance of the LHCb and ALICE
detectors in the fixed target mode is shown in Fig.~\ref{fig:acceptanceComparison-LHC-RHIC}
in comparison with the collider mode and experiments at RHIC, showing clearly the uniqueness of such a setup.

\section{High $x$ frontier}
We review here selected projections of the impact of the fixed target experiment on proton and
nuclear parton distribution functions (PDFs). The projections have been performed for the case
of proton beam colliding either with protons or heavy targets in both cases leading to nucleon-nucleon
c.m.s. energy of $\sqrt{s_{NN}}=115$ GeV. We assume integrated luminosity of 10 fb$^{-1}$ in the $pp$ case
and 100 pb$^{-1}$ in proton-nucleus case (where we considered the Xe and W nuclei).
The projections were done using either the profiling~\cite{Paukkunen:2014zia,Alekhin:2014irh}
or the reweighting method~\cite{Giele:1998gw,Kusina:2016fxy}.

\begin{figure}[th]
\centering{}
\subfloat[$pp$ case]{\includegraphics[width=0.24\textwidth]{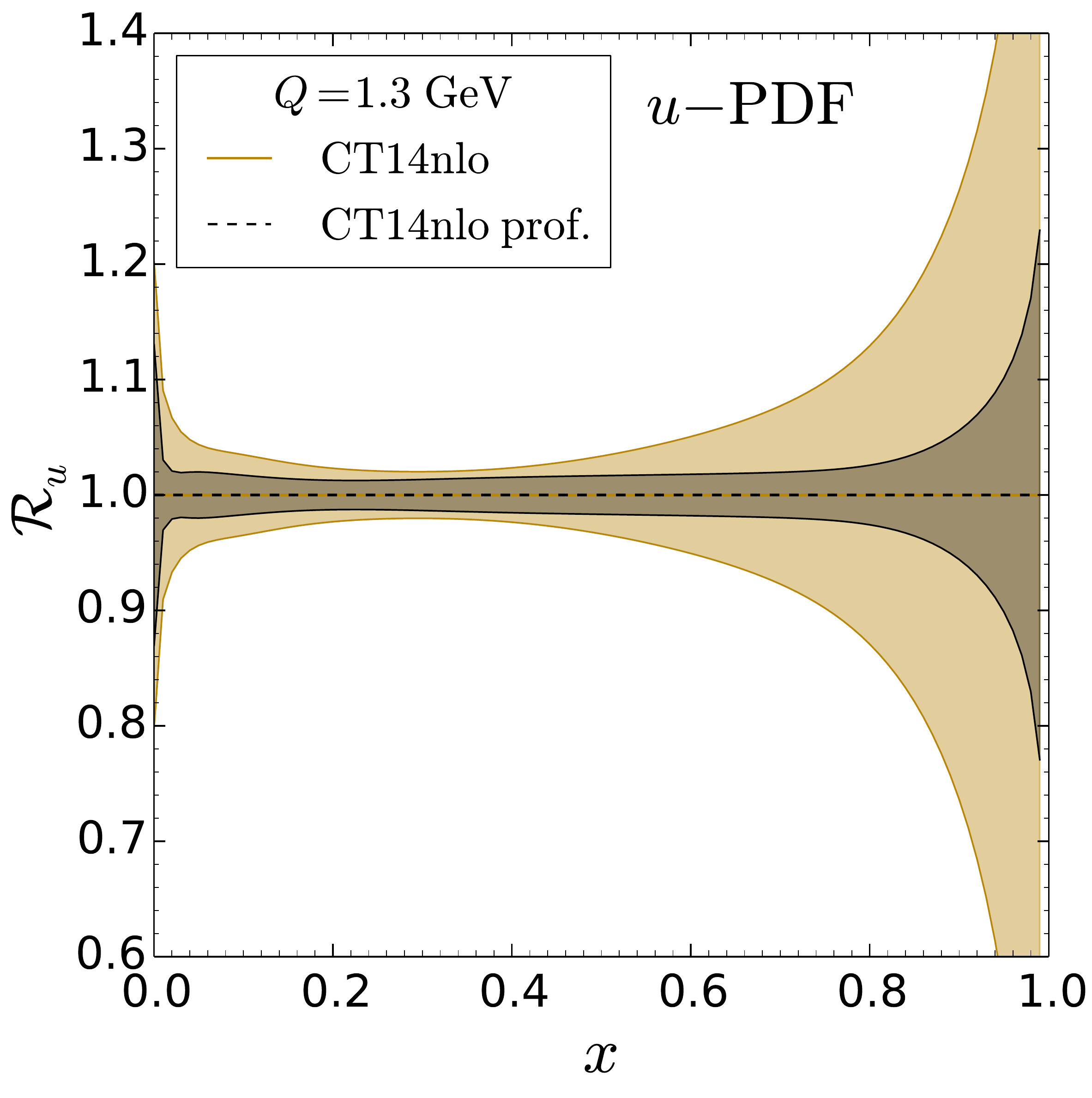}\label{fig:DYpp_a}}
\subfloat[$pp$ case]{\includegraphics[width=0.245\textwidth]{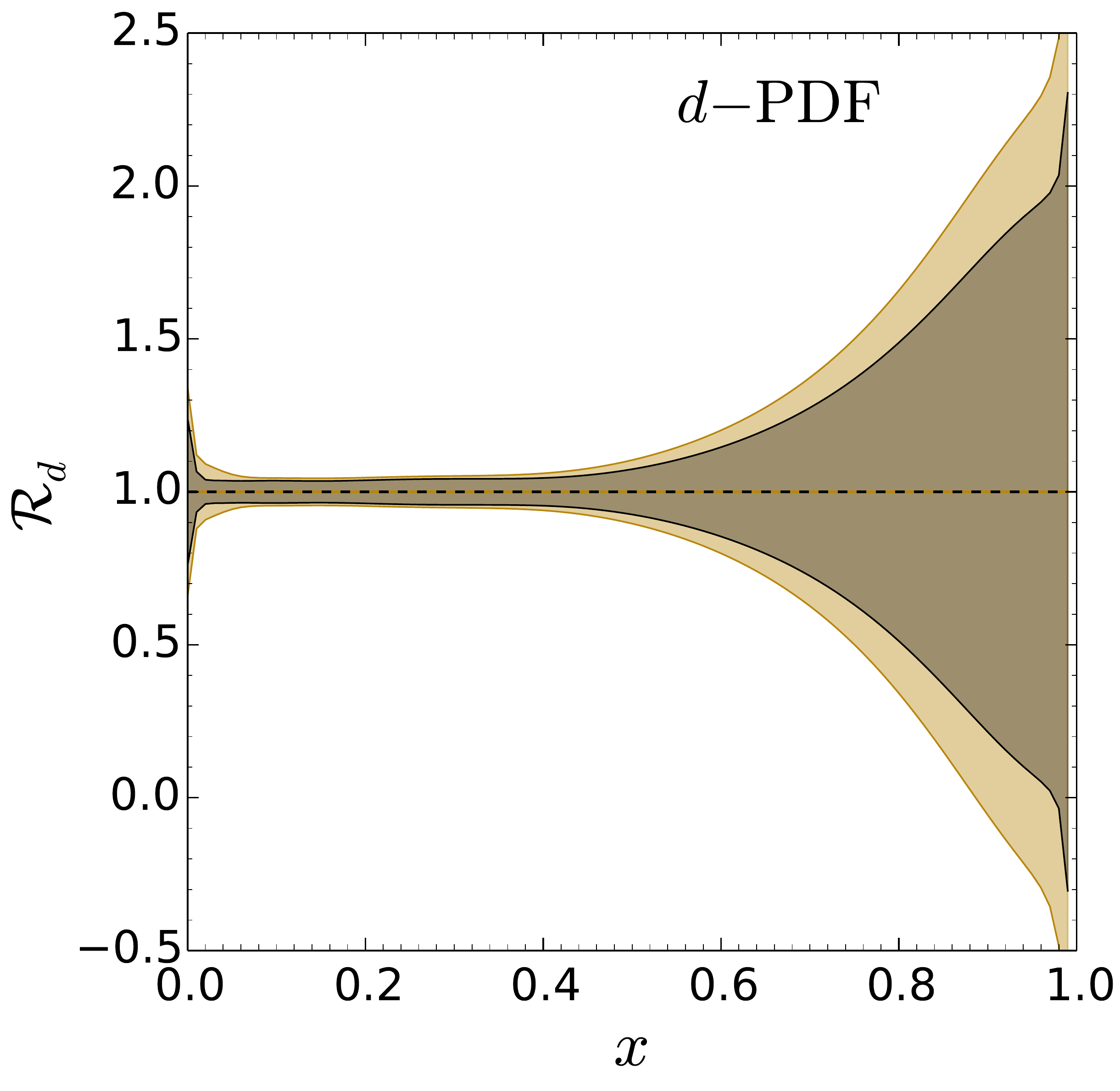}\label{fig:DYpp_b}}
\subfloat[$p$A case]{\includegraphics[width=0.24\textwidth]{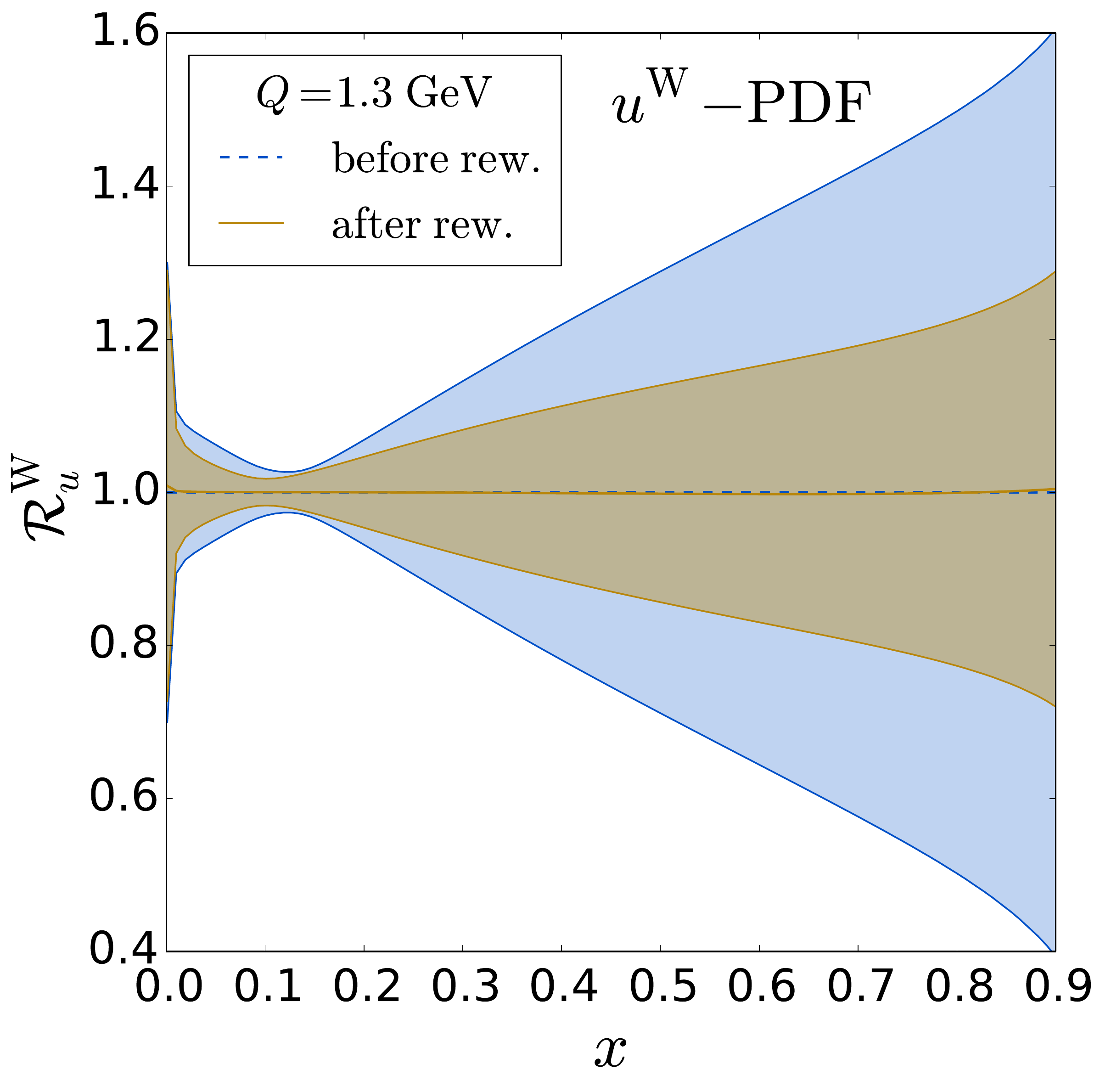}\label{fig:DYpA_a}}
\subfloat[$p$A case]{\includegraphics[width=0.24\textwidth]{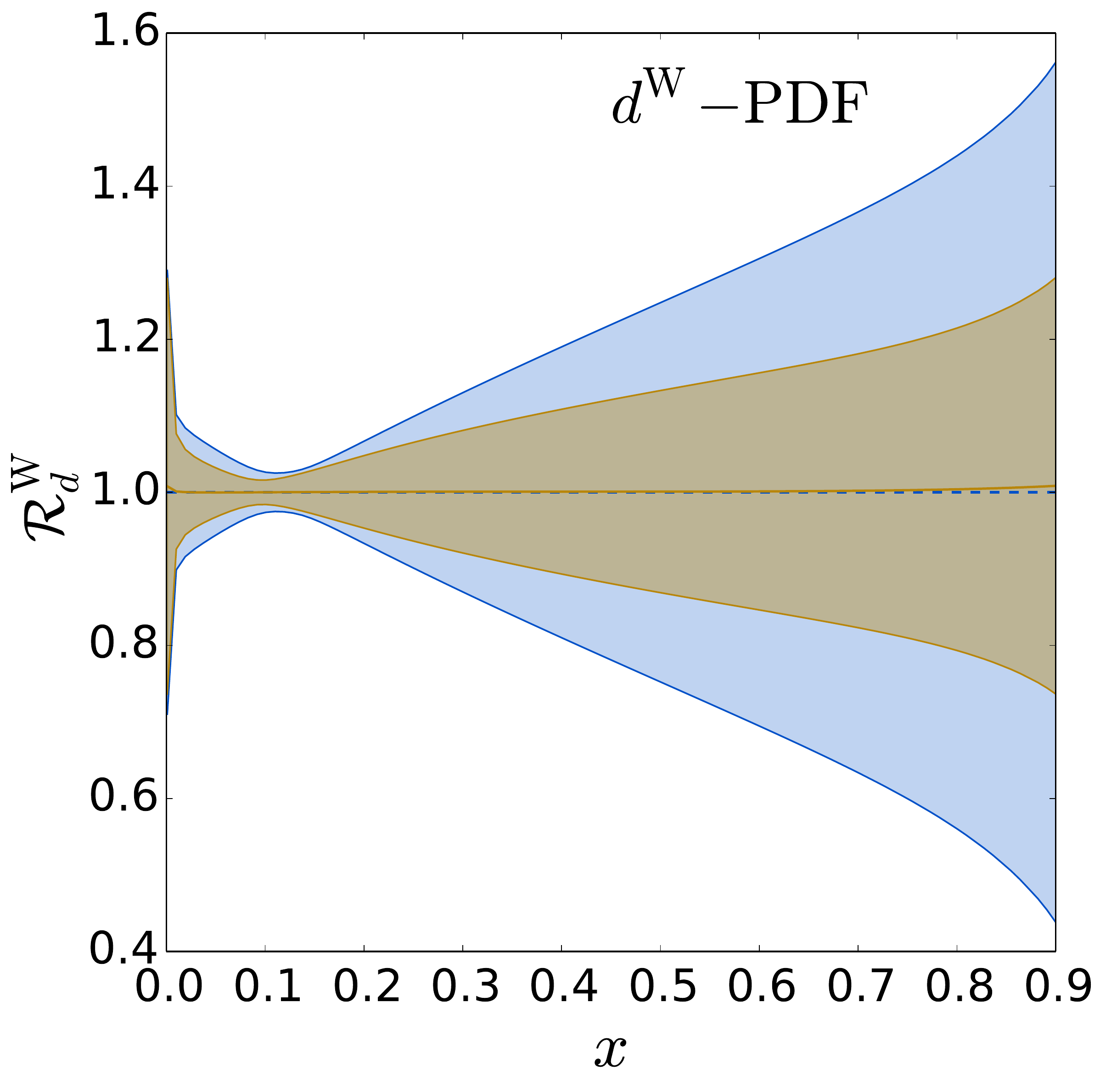}\label{fig:DYpA_b}}
\caption{\small 
(a,b) Impact of the Drell-Yan lepton pair production in $pp$ collisions at $\sqrt{s_{NN}}=115$ GeV
on the PDF uncertainties. The $u$ and $d$ PDFs from CT14~\cite{Dulat:2015mca} are plotted
as a function of $x$ at a scale $Q=1.3$ GeV before and after including AFTER@LHCb
pseudo-data using the profiling method.
(c,d) nCTEQ15 nPDFs~\cite{Kovarik:2015cma} before and after the reweighting using $R^{\text{DY}}_{p\text{W}}$
and $R^{\text{DY}}_{p\text{Xe}}$ AFTER@LHCb pseudo-data. The plots show ratio of nPDFs for tungsten
(W) and the corresponding uncertainties compared to the central value at the scale $Q=1.3$ GeV.
Figure from~\cite{Hadjidakis:2018ifr}.}
\label{fig:DY}
\end{figure}

We start with the Drell-Yan lepton pair production process. In the $pp$ case, pseudo-data for
rapidity distributions in selected bins of the lepton pair invariant mass:
$M_{\mu\mu}\in [4,5], [5,6], [6,7], [7,8]$ GeV and $M_{\mu \mu} > 10.5$ GeV have been used; 
assuming the acceptance of the LHCb detector: $2<\eta_{\mathrm{lab}}<5$ 
(giving $-2.8<y_{\mathrm{c.m.s.}}<0.2$) and $p_T^{\mu}>1.2$ GeV.
The results of using these pseudo-data on the uncertainties of the CT14 PDFs~\cite{Dulat:2015mca}
is presented in Figs.~\ref{fig:DYpp_a} and~\ref{fig:DYpp_b} for the case of $u$ and $d$ distributions.
We can see a huge reduction of uncertainties in the high-$x$ region for the $u$ quark and a moderate
reduction for the $d$ quark; a smaller decrease is also observed for the light sea quarks. Additionally,
a sizable decrease of the PDF uncertainties is also found in the intermediate and small $x$ region
which can be seen e.g. in Fig.~15 in~\cite{Hadjidakis:2018ifr}.

In the nuclear case pseudo-data for the nuclear modifications of rapidity distributions in $p$Xe
and $p$W collisions have been used; also assuming the acceptance of the LHCb detector and using
the same bins in the lepton pair invariant mass. In Figs.~\ref{fig:DYpA_a} and~\ref{fig:DYpA_b}
we present the impact of these pseudo-data on the nCTEQ15~\cite{Kovarik:2015cma} nuclear PDFs (nPDFs)
in case of tungsten for $u$ and $d$ distributions. In both cases a very substantial reduction
of the uncertainties is observed. We should highlight here that in reality the nPDF uncertainties
are typically underestimated and in practice this reduction is even larger.

\begin{figure}[th]
\centering{}
\includegraphics[width=0.24\textwidth]{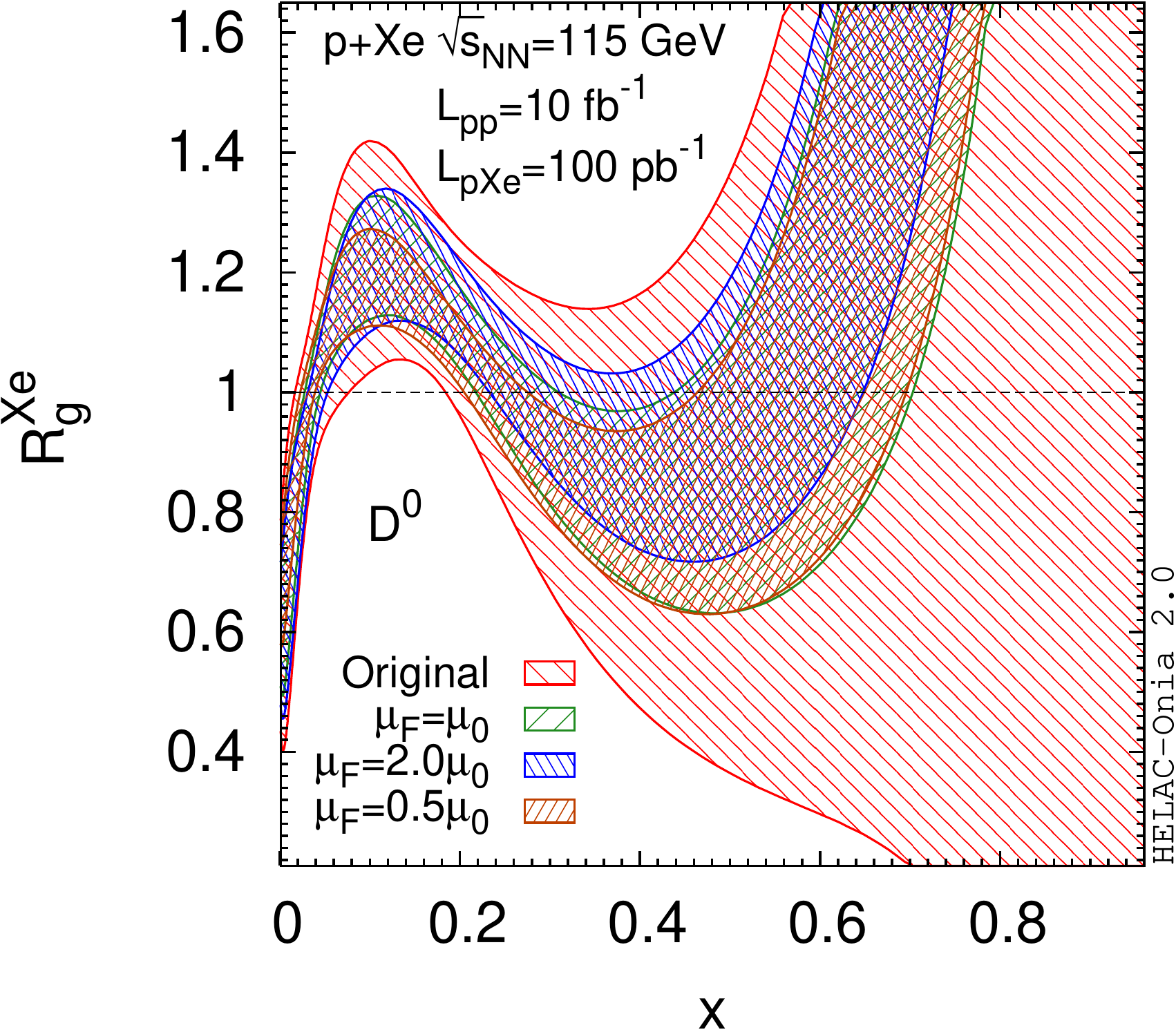}
\includegraphics[width=0.24\textwidth]{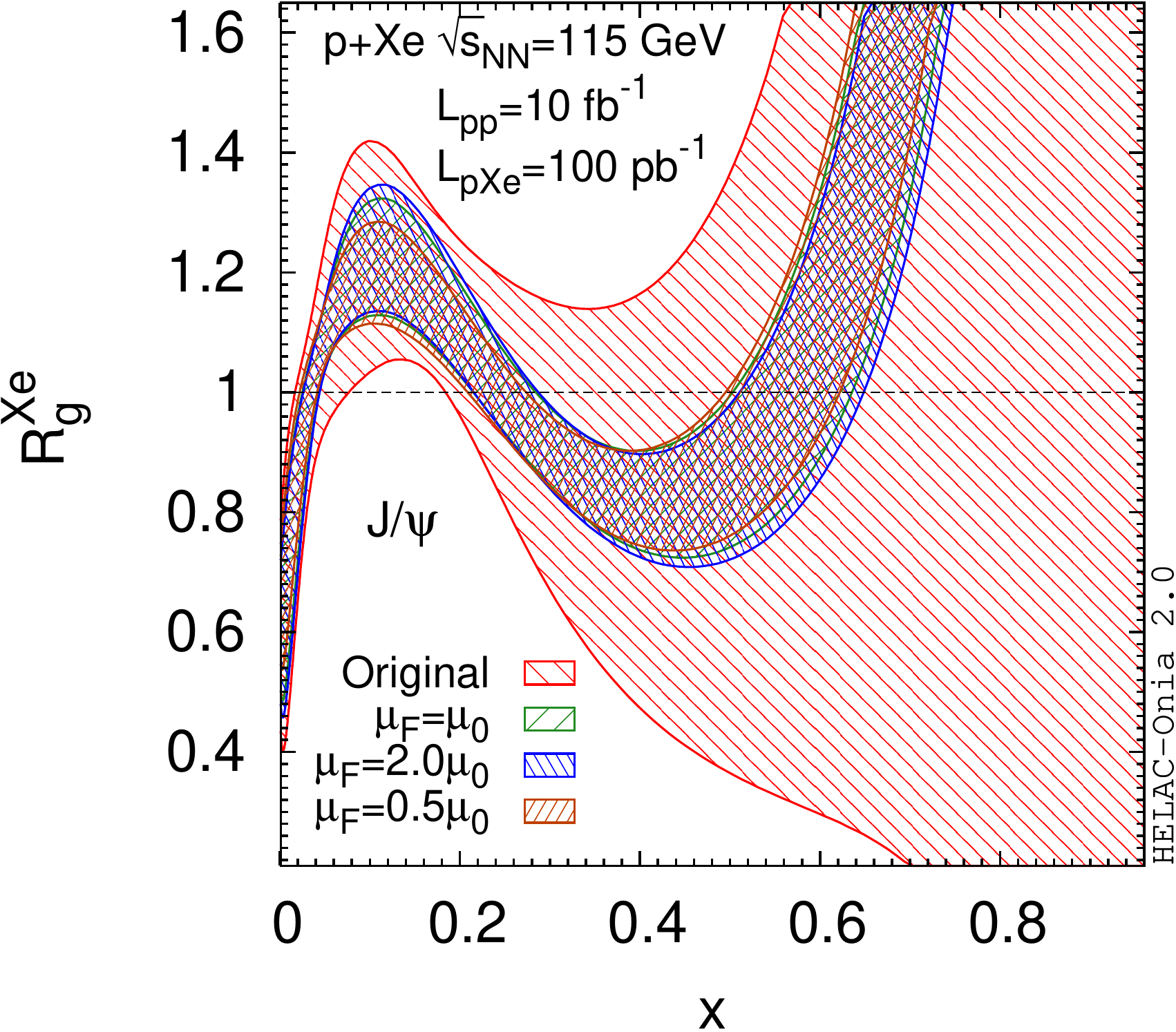}
\includegraphics[width=0.24\textwidth]{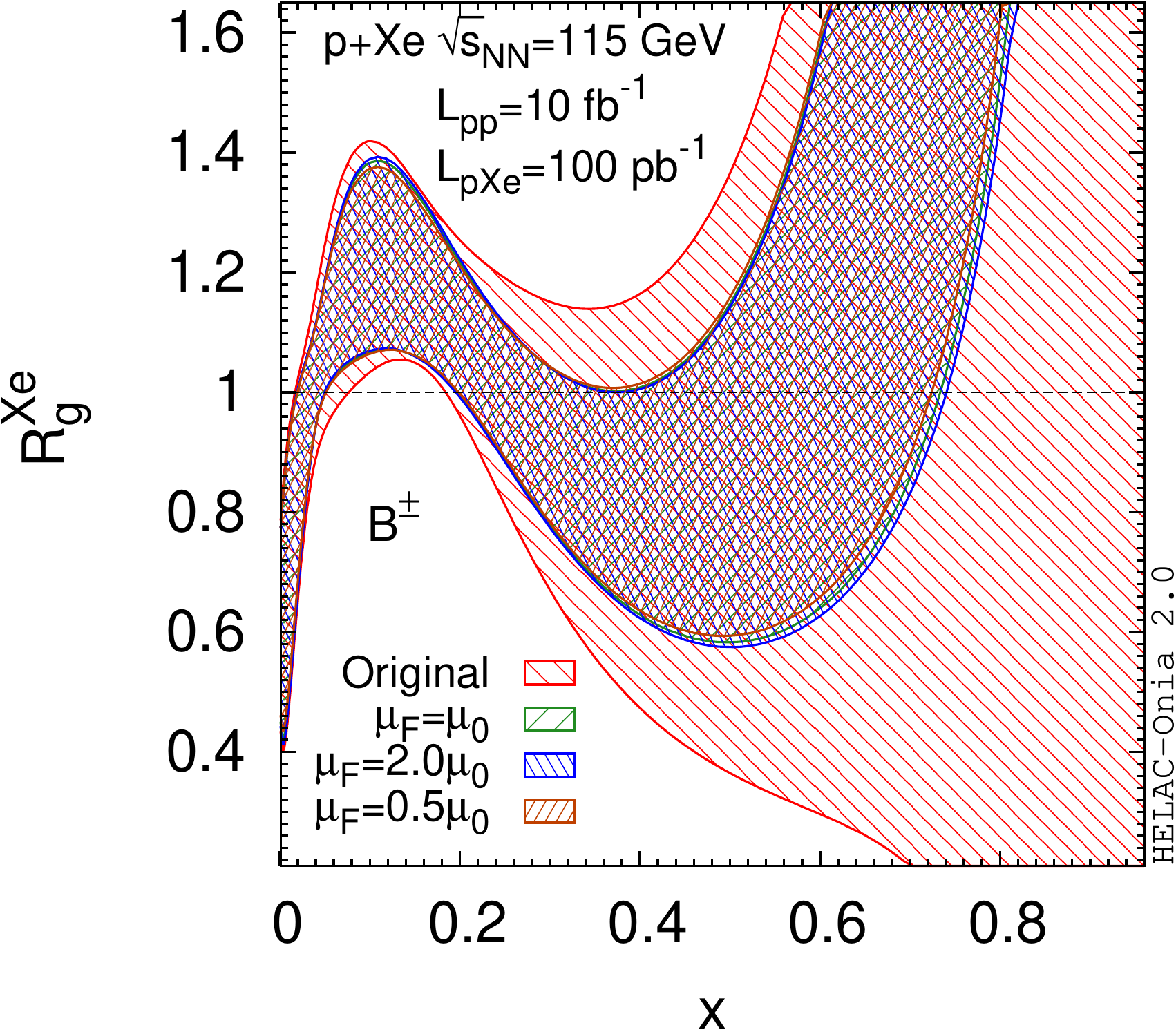}
\includegraphics[width=0.24\textwidth]{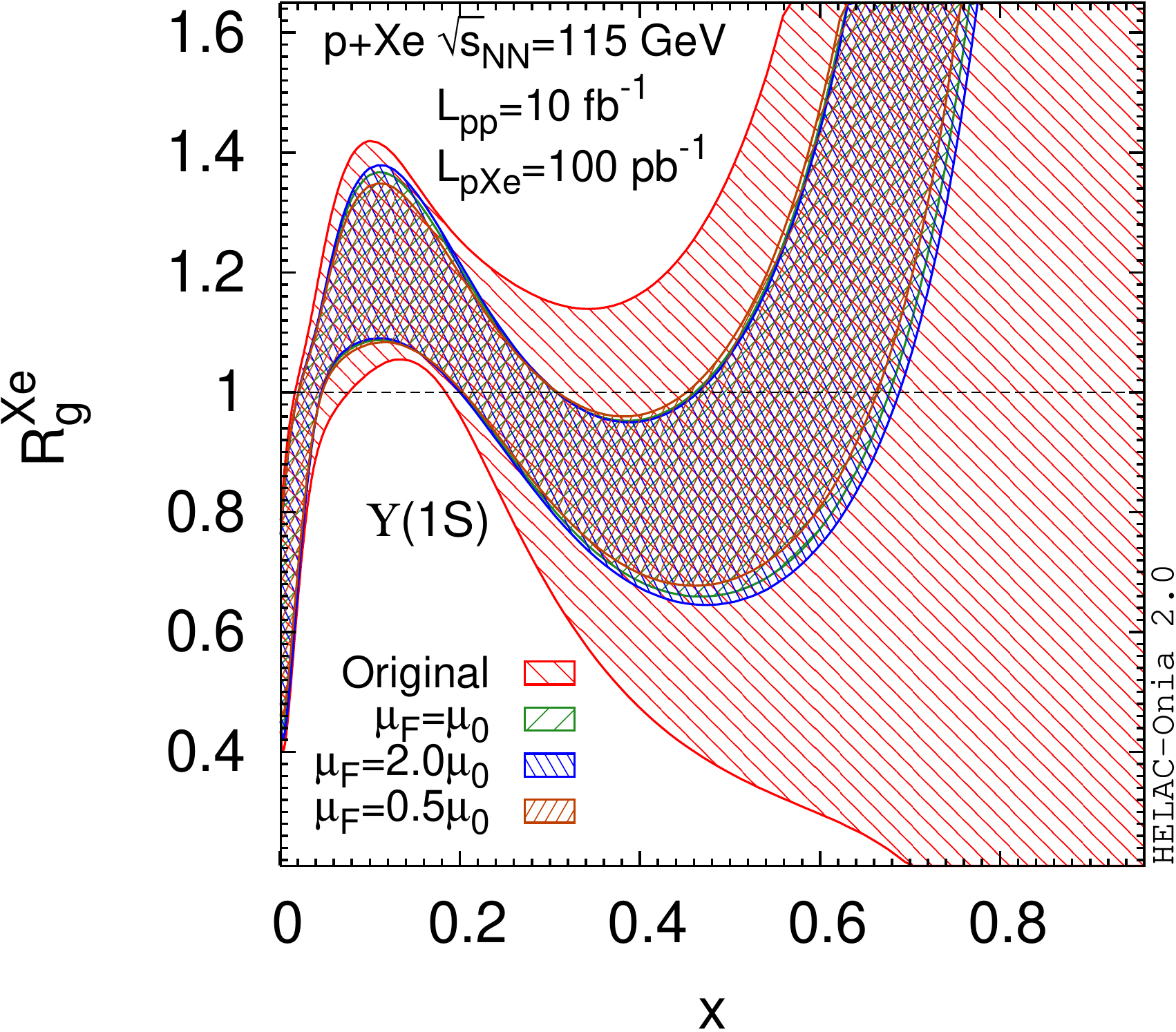}
\caption{\small nCTEQ15 nPDFs before and after the reweighting using $R_{pXe}$ pseudo-data
for (a) $D^0$, (b) $J/\psi$, (c) $B^+$, (d) $\Upsilon(1S)$ production at AFTER@LHCb.
The plots show ratios $R_g^{Xe}$ of gluon densities encoded in nCTEQ15 over that in CT14 PDFs
at scale $Q=2$ GeV. Figure from~\cite{Hadjidakis:2018ifr}.}
\label{fig:gluon}
\end{figure}

As a second example we concentrate on the nuclear gluon distribution which is one of the
least known nPDFs. One example of data that can be used in order to learn about the gluon
nPDF comes from heavy flavour production. The collider LHC data for heavy flavour production in
both $pp$ and $p$Pb collisions have been already used to constrain the small-$x$
gluon~\cite{Zenaiev:2015rfa,Gauld:2016kpd,Kusina:2017gkz},
here we show that the corresponding data collected in the fixed target mode can also be
used to study the high-$x$ gluon distribution.
In Fig.~\ref{fig:gluon} we present results of reweighting of the nCTEQ15 nPDFs using pseudo-data
for the production of heavy flavour mesons ($D^0,J/\psi,B^{\pm},\Upsilon(1S)$) in $p$Xe collisions
at $\sqrt{s_{NN}}=115$ GeV using the HELAC-Onia framework~\cite{Shao:2012iz,Shao:2015vga}.
We can see that in all cases it allows for a huge reduction of the
nPDF uncertainties of the gluon distribution at high-$x$, even taking into account the relatively
large scale uncertainties.

\vspace{-0.2cm}
\begin{figure}[!hb]
\centering
\subfloat[~$2<y_{\rm Lab}<3$]{\includegraphics[width=0.31\textwidth]{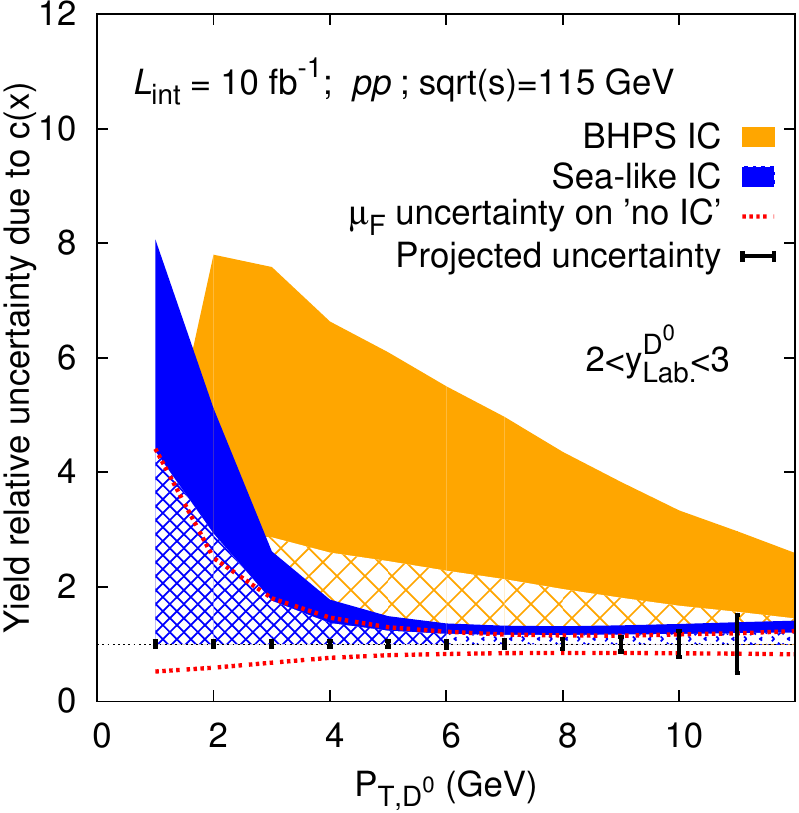}}
\subfloat[~$3<y_{\rm Lab}<4$]{\includegraphics[width=0.31\textwidth]{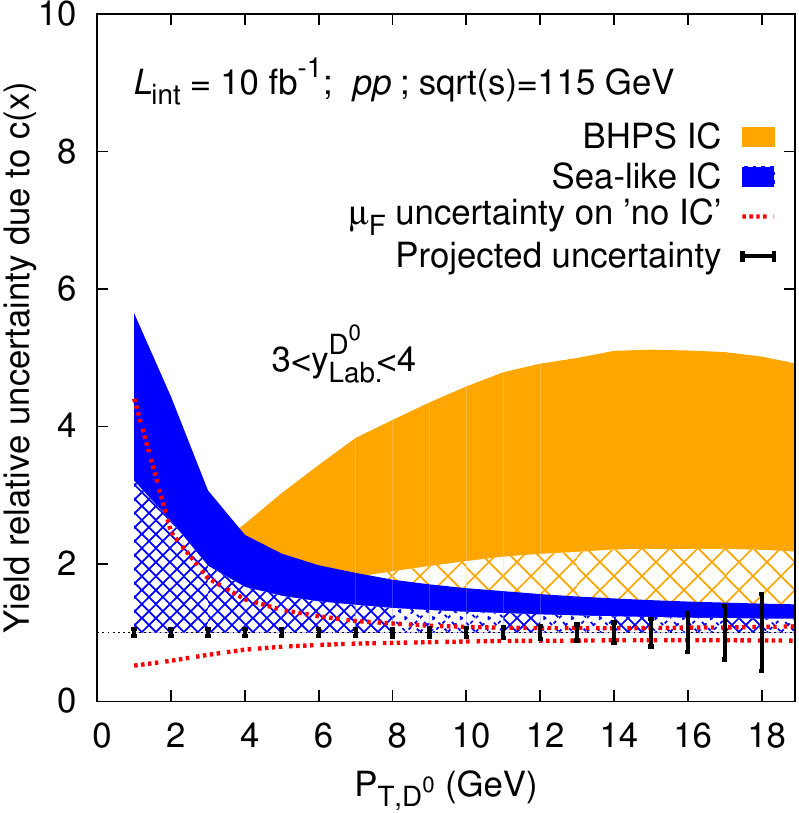}}
\subfloat[~$4<y_{\rm Lab}<5$]{\includegraphics[width=0.31\textwidth]{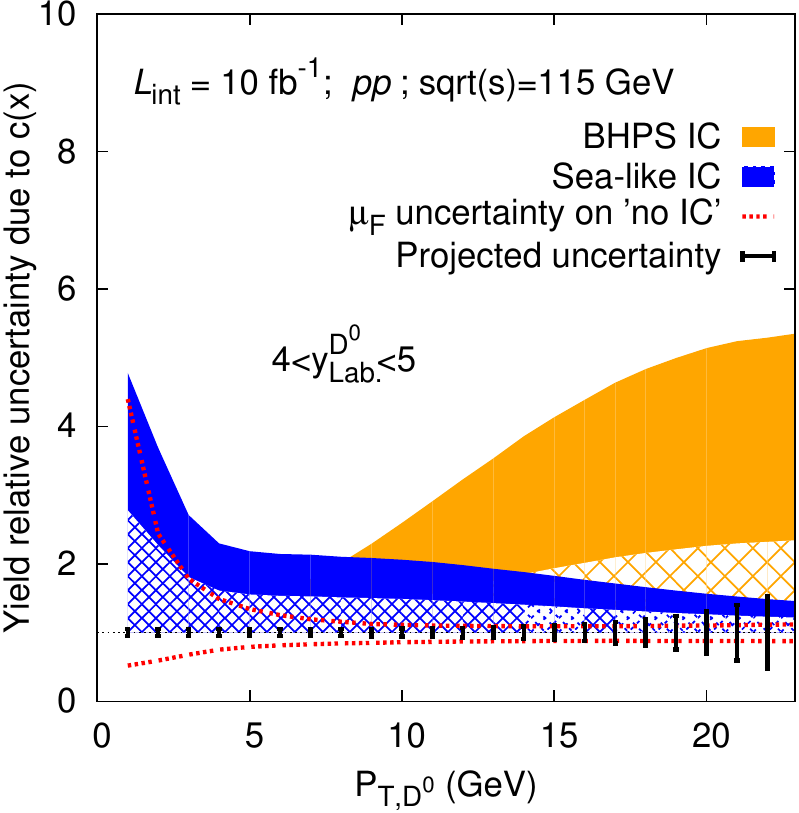}}
\caption{\small 
Impact of the uncertainties on the charm content of the proton on the $D^0$ yield
as a function of $p_T$ compared to the projected uncertainties from the measurement of
the $D^0$ yield in $pp$ collisions at $\sqrt{s_{NN}}=115$ GeV in the LHCb acceptance.
The orange and blue zones correspond to BHPS~\cite{Brodsky:1980pb} and sea-like~\cite{Pumplin:2007wg}
intrinsic charm (IC) models. The filled areas correspond to yields computed with up to
$\langle x_{c \bar c}\rangle= 2\%\, (\text{resp. } 2.4\%)$ and the hashed areas up to
$\langle x_{c +\bar c}\rangle= 0.57\%\, (\text{resp. } 1.1\%)$.
The dashed red lines indicate the factorization scale uncertainty on the ``no-IC'' yield.
Figure from~\cite{Hadjidakis:2018ifr}.}
\label{fig:large_x_2}
\end{figure}
The last example we present in this proceedings is the possible impact of the fixed target
AFTER@LHC experiment on the heavy quark PDFs in the proton. In particular, its possibilities
to confirm or strongly constrain the models of the non-perturbative (intrinsic) charm that
typically predict a substantial component of the heavy quark distribution located at
high-$x$~\cite{Brodsky:2015fna}.
In Fig.~\ref{fig:large_x_2} we show the relative yield uncertainty for inclusive $D^0$ meson
production for three rapidity bins ($2<y_{\rm lab}<3$, $3<y_{\rm lab}<4$, $4<y_{\rm lab}<5$)
as a function of the transverse momentum of the $D^0$ meson.
As can be seen, even for $p_T \lesssim 15$ GeV the expected precision of the measurement will clearly allow to 
considerably constrain the intrinsic-charm model, by up to an order magnitude.

\section{Summary}
We have presented a selection of results from~\cite{Hadjidakis:2018ifr} showing the impact
of a fixed target experiment with LHC beams on our current knowledge of proton and nuclear PDFs.
From these results it is clear that such an experiment would be invaluable in learning about
large-$x$ distributions of quarks and gluons in both proton and nuclei. It would provide a unique
opportunity for studying the high-$x$ regime which is hard to access in a collider mode of
the LHC and at the same time it is important for answering many important questions like:
(i) understanding the confinement properties of QCD,
(ii) understanding the origin of the EMC effect,
(iii) discriminating between different models of hadronic structure,
(iv) testing the existence of intrinsic heavy quarks in the proton,
(v) reducing the uncertainty on prompt neutrino fluxes,
or (vi) improving our knowledge of parton luminosities at existing and future hadron colliders
(LHC, HE-LHC, FCC-hh) which will help with searches of new heavy particles.
The full results of the studies from the AFTER@LHC group can be found in~\cite{Hadjidakis:2018ifr}
in particular topics related to spin physics and heavy ion collisions which were not covered here.

\bibliographystyle{utphys_spires}
\bibliography{refs}

\end{document}